\documentclass[11pt]{article}

\usepackage{graphicx}
\usepackage{multicol}
\usepackage{color}
\usepackage{epsfig,latexsym}
\usepackage{textcomp}

\definecolor{verdon}{cmyk}{1,0.5,1,0}
\definecolor{blue}{cmyk}{0.8,0.8,0,0.}
\definecolor{red}{cmyk}{0.2,1,1,0.0}

\def\lapprox{\mathrel{\mathop  {\hbox{\lower0.5ex\hbox{$\sim$}
\kern-1.1em\lower-0.7ex\hbox{$<$}}}}}
\def\gapprox{\mathrel{\mathop  {\hbox{\lower0.5ex\hbox{$\sim$}
\kern-1.1em\lower-0.7ex\hbox{$>$}}}}}

\begin{document}

\title{\color{verdon} 
ecCNO solar neutrinos: a challenge for gigantic ultra-pure liquid scintillator detectors}

\author{F.L. Villante$^{1,2}$
\vspace{0.5 cm}\\
{\small\em $^{1}$Universit\`a dell'Aquila, Dipartimento di Scienze Fisiche e Chimiche, L'Aquila, Italy}\\
{\small\em $^{2}$INFN, Laboratori Nazionali del Gran Sasso, Assergi (AQ), Italy}}

\date{}

\maketitle

\def\abstractname{\color{red}\bf Abstract}
\begin{abstract}
{\footnotesize 
Neutrinos produced in the Sun by electron capture reactions on $^{13}{\rm N}$,
$^{15}{\rm O}$ and $^{17}{\rm F}$, to which we refer as ecCNO neutrinos, are not usually considered 
in solar neutrino analysis since the expected fluxes are extremely low. 
The experimental determination of this sub-dominant component of the solar neutrino flux
is very difficult but could be rewarding since it provides a determination 
of the metallic content of the solar core and, moreover, probes the solar neutrino survival 
probability in the transition region at $E_\nu\sim 2.5\,{\rm MeV}$. In this letter, we suggest 
that this difficult measure could be at reach for future gigantic ultra-pure liquid scintillator 
detectors, such as LENA.}
\end{abstract}

\newpage

\section{Introduction}

 One of the main goals of the present and next generation ultra-pure liquid scintillator detectors, such as 
Borexino~\cite{BorexinoLong}, SNO+~\cite{SNOPlus} and LENA~\cite{Lena}, 
is the determination of the neutrino fluxes produced by the CNO cycle in the Sun. 
The evaluation of CNO cycle efficiency is, in fact, connected with various important problems, like e.g. 
the determination of globular clusters age \cite{GC} from which we extract a lower limit to the age of the Universe.
Moreover, it can provide clues to solve the so-called ``solar composition problem",
i.e. the fact that Standard Solar Models (SSM) implementing the latest photospheric heavy element abundances~\cite{AGSS09} 
are not able to reproduce the helioseismic results, see e.g. \cite{SerenelliRev,Basu,noi}.

 The present experimental efforts in this direction are devoted to the observation of neutrinos 
originating from the $\beta^+$ decay of $^{13}{\rm N}$, $^{15}{\rm O}$ and $^{17}{\rm F}$, the so-called {\em CNO neutrinos}, 
that represent about $1\%$ of the total solar neutrino budget. Their detection is, however, a
difficult task. The CNO neutrinos have continuous energy spectra with endpoints at about 1.5~MeV 
and do not produce specific spectral features that permit to extract the signal unambiguously from 
the background in high purity liquid scintillators (see e.g. \cite{PoBi} for a discussion).

 In this work, we consider a different source of neutrinos in the CNO cycle that is generally neglected. 
It was pointed out by \cite{ecCNORob} and \cite{ecCNOBah} that neutrinos can be also 
produced in the Sun by electron capture reactions on $^{13}{\rm N}$, $^{15}{\rm O}$ and $^{17}{\rm F}$. 
The resulting fluxes,  to which we refer as {\em ecCNO neutrino} fluxes, are extremely small, at the level 
of $0.1\%$ with respect to the ``conventional" CNO neutrino fluxes. However, ecCNO neutrinos 
are monochromatic and have larger energies equal to $E_\nu\sim 2.5 \, {\rm MeV}$. 

 We suggest that these characteristics, together with the development of gigantic (i.e. with masses $\sim 10\; {\rm kton}$ or more) 
ultra-pure liquid scintillator detectors, such as LENA \cite{Lena}, could make their detection possible. 
Clearly, the determination of this sub-dominant component of the solar neutrino flux is extremely difficult but could 
be rewarding in terms of physical implications. In fact, besides testing the efficiency of the CNO cycle, 
ecCNO neutrinos could permit to determine the metallic content of  the solar core and also probe 
the electron neutrino survival probability in an energy region that is otherwise inaccessible, with important 
implications for the final confirmation of the LMA-MSW flavour oscillation paradigm.

 Is thus useful to investigate the potential of future gigantic liquid scintillator experiments 
for ecCNO neutrino detection. With this spirit, in sect.~\ref{ecCNO}  we review and update the predictions 
of ecCNO neutrinos fluxes by \cite{ecCNORob} in light of the recent SSM calculations 
and we calculate the event spectrum expected in liquid scintillator experiments. 
In sect.~\ref{background}, we compare our results with the expected background rates. 
In sect.~\ref{summary}, we give our conclusions.

\begin{table}[t]
\begin{center}{
\begin{tabular}{l |cc  }
      Fluxes    &    GS98       & AGSS09         \\
\hline
\hline 
      $\Phi_{\rm N}$       &    $2.96\,(1 \pm 0.14)\times 10^8$       &   $2.17\,(1 \pm 0.14)\times 10^8$ \\
      $\Phi_{\rm O}$    &      $2.23\,(1 \pm 0.15)\times 10^8$     &    $1.56\,(1 \pm 0.15)\times 10^8$ \\
      $\Phi_{\rm F}$    &       $5.52\,(1 \pm 0.17)\times 10^6$    &    $3.40\,(1 \pm 0.17)\times 10^6$ \\               
\hline
     $\Phi_{\rm eN}$    &      $2.34\,(1 \pm 0.14)\times 10^5$     &  $1.71\,(1 \pm 0.14)\times 10^5$ \\ 
    $\Phi_{\rm eO}$      &      $0.88\,(1 \pm 0.15)\times 10^5$     &  $0.62\,(1 \pm 0.15)\times 10^5$ \\
   $\Phi_{\rm eF}$       &     $3.24\,(1 \pm 0.17)\times 10^3$      &  $2.00\,(1 \pm 0.17)\times 10^3$ \\ 
\hline
$\Phi_{\rm B}$ & $5.58\,(1 \pm 0.14)\times 10^6$   & $4.59\,(1 \pm 0.14)\times 10^6$  \\
\hline
\end{tabular}
}\end{center}\vspace{0.4cm} \caption{\em {\protect\small . The neutrino fluxes (${\rm cm}^{-2}\,{\rm s}^{-1}$)
produced by $\beta^+$ decay and electron capture processes on  $^{13}{\rm N}$, $^{15}{\rm O}$ and $^{17}{\rm F}$ nuclei in the sun.
In the last line, we also give the predictions for the $^{8}{\rm B}$ neutrino flux. 
\label{tab1} 
}}\vspace{0.4cm}
\end{table}

\section{ecCNO neutrinos}
\label{ecCNO}

 The dominant hydrogen burning mechanism in the Sun is the pp-chain
which accounts for $\sim 99\%$ of the total energy (and neutrino) production. 
A sub-dominant contribution is given by the CNO cycle 
that produces significant neutrino fluxes originating from the $\beta^+$ 
decay of $^{13}{\rm N}$, $^{15}{\rm O}$ and $^{17}{\rm F}$. The SSM predictions for the CNO neutrino fluxes 
are given in the first three lines of Tab.\ref{tab1}. These values were obtained in \cite{serenelli2011}
by assuming, as input for SSM calculations, 
the ``old'' high surface metallicity of GS98 \cite{GS98} 
and the ``new''  low surface metallicity of AGSS09 \cite{AGSS09}.

 As it was pointed out in \cite{ecCNORob,ecCNOBah}, along with
these fluxes, neutrinos are produced by the electron capture reactions:
\begin{eqnarray}
\nonumber
^{13}{\rm N}+e^-&\rightarrow&^{13}{\rm C}+\nu_e\\
\nonumber
^{15}{\rm O}+e^-&\rightarrow&^{15}{\rm N}+\nu_e\\
^{17}{\rm F}+e^-&\rightarrow&^{17}{\rm O}+\nu_e
\end{eqnarray}
These neutrinos, which we refer to as ecCNO neutrinos, are monochromatic with energies 1.022 MeV above the $\beta^+$
spectrum endpoints that correspond to $E_{\nu}= 2.220$, $2.754$,  and $2.761$~MeV, respectively.

 We briefly review the calculation of the ecCNO neutrino fluxes performed by \cite{ecCNORob} in order to
obtain updated predictions that take into account the recent revisions in SSMs calculations. 
The ratios $r$ between $\beta^+$ decay and electron capture rates for $^{13}{\rm N}$, $^{15}{\rm O}$ and $^{17}{\rm F}$ nuclei
are measured {\em in laboratory} and are given by $r= 1.96 \times 10^{-3}$, $9.94\times10^{-4}$ 
and $1.45\times 10^{-3}$ \cite{ecCNORob}, respectively. 
However, the electron capture rates {\em in the sun} have to be rescaled proportionally 
to the electron number density at the nuclear site which has to be calculated by 
taking into account: the distortion of the electron wave functions in the Coulomb field
of nuclei; electron capture from bound states; screening effects, as it is e.g. discussed
in \cite{GruzBah} where a comprehensive analysis of the $^7{\rm Be}$ electron capture was
given. The ratios $\tilde{r}$ between electron capture rates in the Sun and laboratory are calculated in 
\cite{ecCNORob} by using the temperature and electron density profile of the Sun predicted by
\cite{OldSSM}. The values obtained by averaging over the entire solar 
volume are $\tilde{r}=0.403$, $0.398$, and $0.405$, respectively. One calculates then the 
ecCNO neutrino fluxes by using $\Phi_{\rm eX} = r\times \tilde{r} \times\Phi_{\rm X}$, where $\Phi_{\rm eX}$ ($\Phi_{\rm X}$)
is the flux produced by the electron capture ($\beta$ decay) of the ${\rm X}$ nucleus in the Sun and 
${\rm X}= {\rm N,\; O\, and \;  F}$; the results 
are shown in Tab.~\ref{tab1}. The differences between the quoted values 
and those obtained in \cite{ecCNORob} are due to the fact that recent SSM calculations 
predict smaller ``conventional'' CNO fluxes, 
mainly as a consequence of revised solar surface composition
and updated $S_{1,14}$ astrophysical factor \cite{SolFus}.
Note that we made the reasonable assumption that the small differences in the temperature profile
of the Sun which are implied by a different choice of the surface composition
do not affect  the $\tilde{r}$ parameter in a significant way. In this assumption, 
the ecCNO neutrinos carry exactly the same information as CNO neutrinos on the 
efficiency of the CNO cycle and on the metallic content of the solar core. 
They probe, however, the solar neutrino survival probability at a different energy, $E_\nu\sim 2.5\,{\rm MeV}$,
which well corresponds to the transition between vacuum averaged and matter enhanced 
neutrino oscillations.

\begin{figure}[t]
\par
\begin{center}
\includegraphics[width=8cm,angle=0]{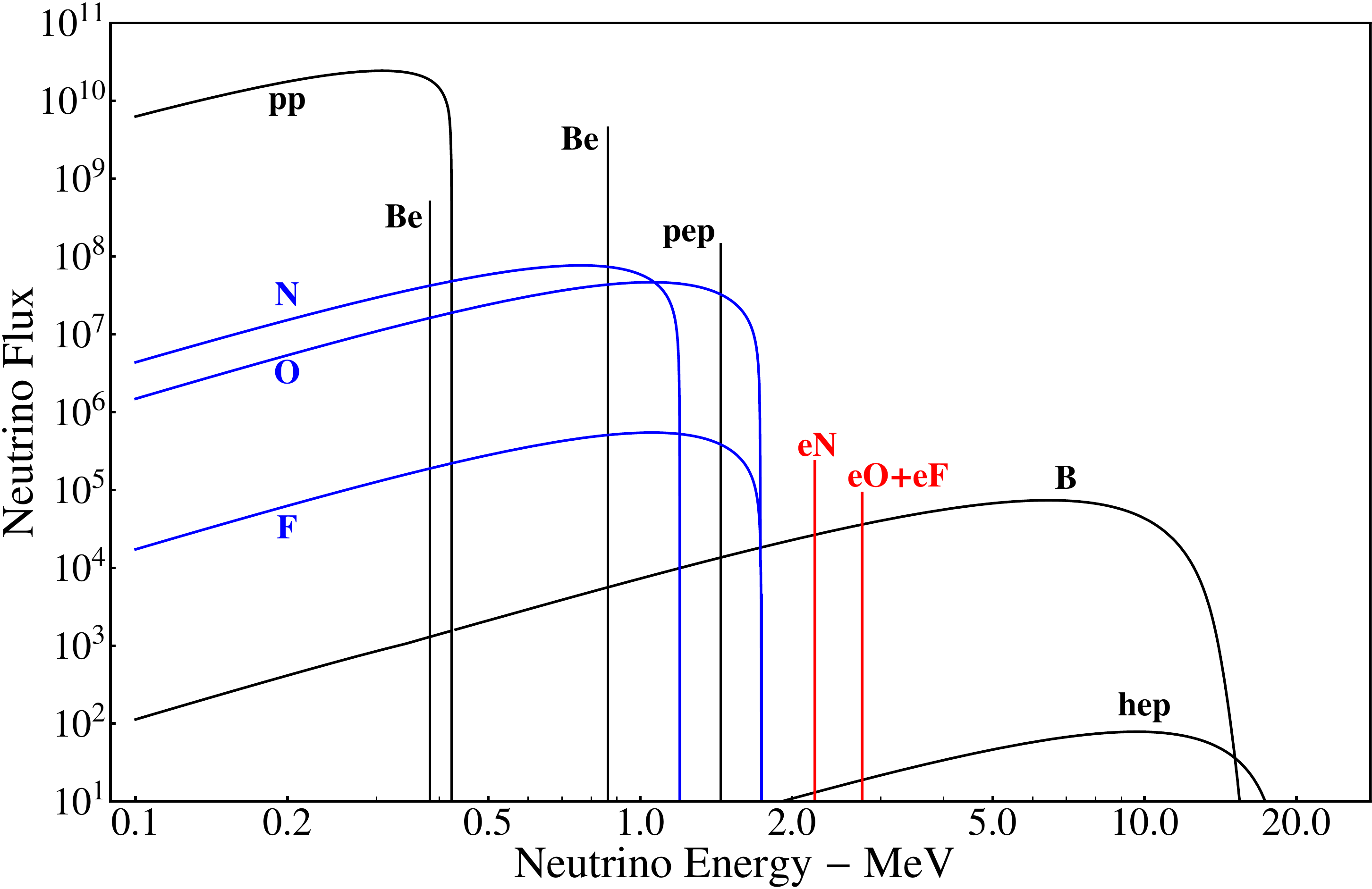}
\includegraphics[width=8cm,angle=0]{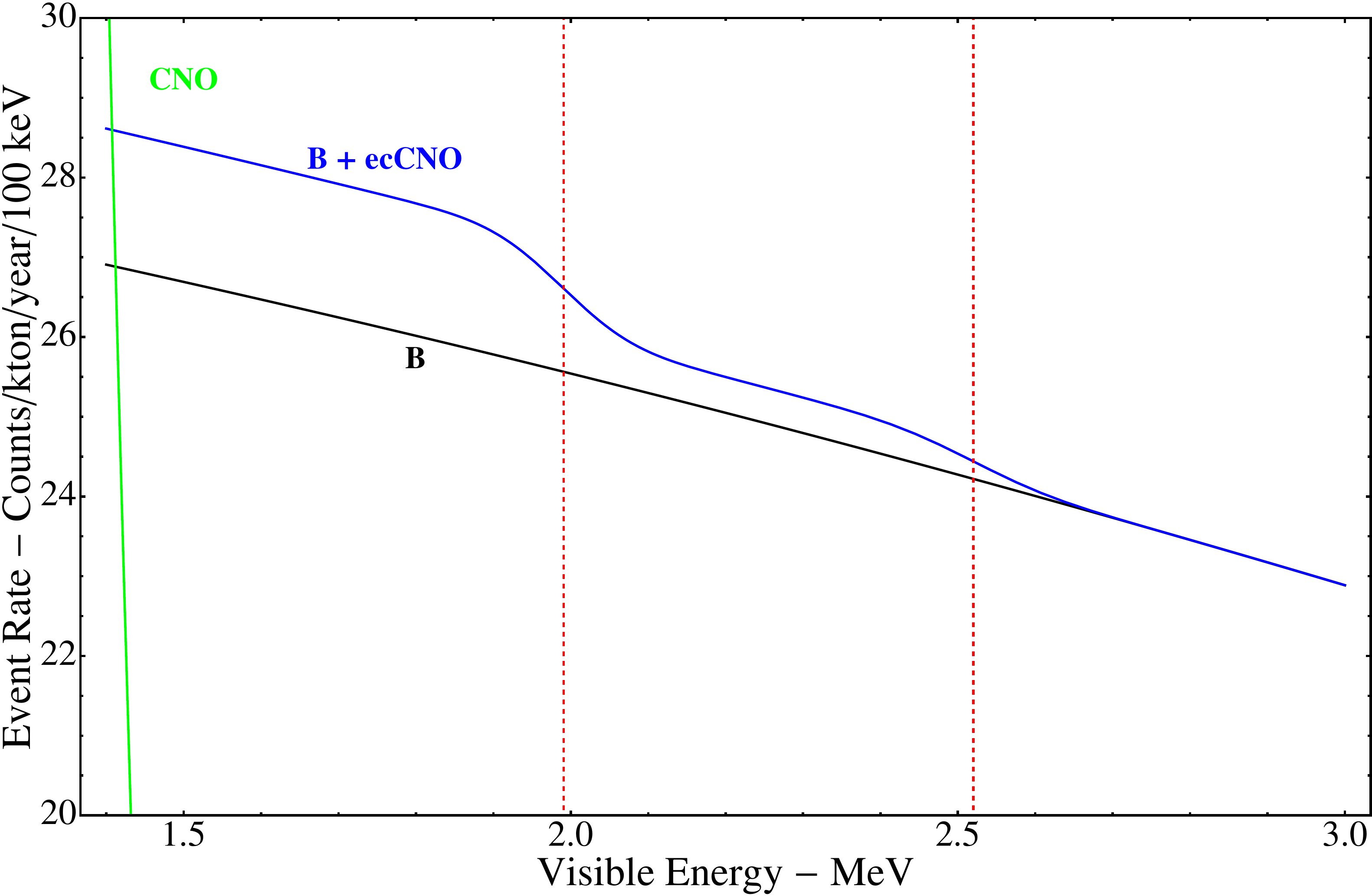}
\end{center}
\par
\vspace{-0.5cm}
\caption{\em\protect\small {\rm Left Panel:} The solar neutrino spectrum; {\rm Right Panel:} The event rate produced by solar neutrinos in liquid scintillator detectors.}
\vspace{0.5cm} 
\label{Fig1}
\end{figure}

 The spectrum of solar neutrinos, including the ecCNO neutrinos
contribution, is shown in the left panel of Fig.~\ref{Fig1} which updates the original figure 
produced by \cite{ecCNORob}. The continuous fluxes are given in ${\rm cm}^{-2} \cdot {\rm s}^{-1}\cdot \left({\rm 100 \, keV}\right)^{-1}$
while the monochromatic fluxes are given in ${\rm cm}^{-2}\cdot {\rm s}^{-1}$.
The eF and eO component have been summed since their energies are almost equal; 
the eF contribution is, however, largely sub-dominant and will be neglected in the following.
At the ecCNO neutrinos energies, the low energy tail of the $^{8}{\rm B}$ neutrino spectrum is also produced.
The figure shows that ecCNO neutrinos emerge over the $^{8}{\rm B}$ contribution 
if they are observed in an hypothetical detector with a spectral response 
as narrow as $\sim {\rm few} \times 100\,{\rm keV}$ or better. 
In this respect, the optimal detector has an 
energy resolution 
at the $10\%$ level or better
and is based on a detection reaction 
that does not wash-out the information on the incoming neutrino energy 
(like e.g. charged current reaction on nuclei). 
Liquid scintillators meet the energy resolution requirement but, unfortunately, 
are based on a detection process ($\nu-e$ elastic scattering) that
provides a response proportional to the integrated flux above the observation energy\footnote{
An interesting option for ecCNO neutrino detection could arise, in the future, 
from ${\rm Li}$-doped water-based liquid scintillators (see  e.g. \cite{Orebi}  
where they are proposed as a target for solar neutrinos) since these detectors could 
combine the requirement of good energy resolution with a detection reaction 
with good spectral response (i.e. charged current reaction on $^{7}{\rm Li}$). 
An investigation of this possibility is outside the goals of this work and will be considered elsewhere.}. 
The integrated $^{8}{\rm B}$ neutrino flux is a factor $\sim 20$ larger than the expected ecCNO neutrino
fluxes and it represents an irreducible background from which the ecCNO neutrino signal have to 
be extracted statistically.
In the last line of tab.~\ref{tab1}, we also report the SSMs predictions for the $^{8}{\rm B}$ neutrino flux.

In the right panel of Fig.~\ref{Fig1}, we show the event rate produced by solar neutrinos through $\nu-e$
elastic scattering in liquid scintillator detectors
in the visible energy region between $E_{\rm vis} = 1.4-3.0\;{\rm MeV}$. 
We assume that the scintillator is based on 
linear-akyl-benzene (LAB) which corresponds to that used in 
the future SNO+ and LENA detectors. 
 We include the effect of LMA-MSW flavour oscillations 
by using the electron neutrino survival probability $P_{\rm ee}$ of \cite{pdg}. This corresponds 
to  $P_{\rm ee}\simeq 0.48$ and $\simeq 0.46$ for eN and eO neutrinos respectively, 
and to $\langle P_{\rm ee} \rangle \simeq 0.37$ when averaged over the entire $^{8}{\rm B}$
neutrino spectrum. We consider the neutrino fluxes predicted by SSMs
that implements the GS98 admixture since these models produce a much better 
description of helioseismic observables \cite{noi}. Finally, we describe the detector energy resolution 
by a Gaussian with an energy dependent width that scale as $5\% \cdot \sqrt{E_{\rm ev}/{\rm MeV}}$, 
where $E_{\rm ev}$ is the average energy of the event. 

 Being mono-energetic, eN and eO neutrinos
produce Compton-shoulders, smeared by resolution effects, at the energies $E_{\rm vis} = 1.99\,{\rm MeV}$ and $2.52\,{\rm MeV}$ 
indicated by the two vertical dashed lines in Fig.~\ref{Fig1}. These shoulders can be identified 
if the detector counting rate is sufficiently high. 
The energy integrated rates produced by ecCNO neutrinos are given by:
\begin{eqnarray}
\nonumber
R^{\rm tot}_{\rm eN} &=& 26 \; {\rm counts}/{\rm1  kton}/{\rm year} \\
R^{\rm tot}_{\rm eO} &=& 12 \; {\rm counts}/{\rm1  kton}/{\rm year} 
\end{eqnarray} 
We expect, however, that ecCNO neutrino signal is unobservable below $E_{\rm vis}\simeq 1.5\,{\rm MeV}$
due to the much larger contribution provided by the conventional CNO neutrinos shown by the green line in
the right panel of Fig.\ref{Fig1}.
In order to explore the possibility to extract the ecCNO neutrino signal, we define the observation
window $[E_{\rm low}, \; 2.5\,{\rm MeV}]$ and we calculate the rates for ecCNO and $^{8}{\rm B}$ neutrinos 
as a function of $E_{\rm low}$ in Tab.~\ref{tab2}.
We estimate the significativity $\Sigma$ of a possible measure by comparing the 
expected signal ${\mathcal S}$ due to ecCNO neutrinos 
to the statistical fluctuations of the 
background ${\mathcal B}$ produced by $^8{\rm B}$ neutrinos. We obtain:
\begin{equation}
\Sigma = \frac{\mathcal S}{\sqrt{\mathcal B}} = 
\frac{\left(R_{\rm eN}+R_{\rm eO}\right)}{\sqrt{R_{\rm B}}} \, \sqrt{\mathcal E} =
\eta \, \sqrt{\mathcal E / ({\rm10  kton}\times{\rm year})}
\label{sigma}
\end{equation}
where ${\mathcal E}$ is the assumed detector exposure. 
In the above formula, since the signal has to be extracted from observed rate by subtraction, 
one implicitly assumes that the background in $[E_{\rm low}, \; 2.5\,{\rm MeV}]$ is known 
from independent observations with a fractional uncertainty lower than 
$1/\sqrt{\mathcal B}\sim {\rm few}\, \%$. 
This is plausible, from a statistical point of view. Indeed, $^8{\rm B}$ solar 
neutrinos and cosmogenic $^{11}{\rm C}$ nuclei, which are the dominant background sources (see below), 
mostly produce events outside the observation window $[E_{\rm low},\,2.5\,{\rm MeV}]$
and can thus be well constrained from observations in other spectral regions. 
This clearly requires that systematical errors on the spectral shapes (at few~MeV) 
of $^8{\rm B}$ and $^{11}{\rm C}$ background are at the $\%$ level. 
 
The parameter $\eta$ in eq.~(\ref{sigma}) represents the statistical significance of a measure with an exposure 
${\mathcal E}= {\rm10 \,  kton}\times{\rm year}$
and is given in the last column of  tab.~\ref{tab2}.
We understand that detectors with fiducial masses equal to $\sim 10\; {\rm kton}$ or more
are necessary, for statistical reasons, to extract the ecCNO neutrino signal. 

\begin{table}[t]
\begin{center}{
\begin{tabular}{l|ccc| c}
 $E_{\rm low} ({\rm MeV})$  &   $R_{\rm eN} $       & $R_{\rm eO} $     &  $R_{\rm B}$      & $\eta $    \\
\hline 
1.5 & 5.9 & 4.4  &  255 &   2.1 \\
1.6 & 4.7  & 4.0  &  229  &   1.8 \\
1.7 & 3.5 &  3.5 &  202 &   1.6 \\
1.8 & 2.3  & 3.0  &   176 &   1.3 \\
1.9 &  1.1&  2.6  &  150  &   1.0 \\
2.0 &  0.3 &  2.1  &  125  &   0.7 \\
\hline
\end{tabular}
}\end{center}\vspace{0.4cm} \caption{\em {\protect\small 
The event rates (${\rm counts}/{\rm1  kton}/{\rm year}$) produced by ecCNO and $^{8}{\rm B}$ neutrinos in the 
energy window $[E_{\rm low},\, 2.5 {\rm MeV}]$. The last column give the values of the parameter $\eta$ defined in
eq.~(\ref{sigma}).
\label{tab2} 
}}\vspace{0.4cm}
\end{table}

\section{Background}
\label{background}

 There are three additional types of background for the proposed measure:
{\em i)} external gamma rays emitted by the materials that contain and surround 
the scintillator; 
{\em ii)} intrinsic radioactive background; 
{\em iii)} cosmogenic radio-isotopes produced
in the liquid scintillator by traversing muons.

 The external gamma background, which presently prevents a measurement of the 
solar $^8{\rm B}$ spectrum below 3~MeV in Borexino can be suppressed by self-shielding.
It was shown in \cite{Lena8B} that
this background source can be reduced at a negligible level
in the proposed $50~{\rm kton}$ LENA detector by applying 
a stringent volume cut that reduce the fiducial mass to $19\,{\rm kton}$. 
This mass would be still sufficient for the proposed measure. 

 The intrinsic background depends on the radio-purity levels that will be achieved
in the detector. We take as a reference the contamination levels that  
were obtained by Borexino during Phase-I \cite{BorexinoLong}\footnote{
Borexino Phase-II reduced by an additional factor $\sim 10$ the 
$^{238}{\rm U}$ and $^{232}{\rm Th}$ contamination levels \cite{BorexinoPP}.}.
In the $^{238}{\rm U}$ chain, the radioisotope that produces events 
in the considered energy window is $^{214}{\rm Bi}$ that undergoes 
$\beta$-decay to $^{214}{\rm Po}$ with a total energy release $Q=3.3\, {\rm MeV}$. 
The $^{238}{\rm U}$  contamination in Borexino is $(5.3\pm0.5)\times 10^{-18} \,{\rm g/g}$ 
and corresponds to a total rate 
$R^{\rm tot}(^{214}{\rm Bi})\simeq 2\times 10^3 {\rm counts}/{\rm year}/ {\rm kton}$.
Fortunately, $^{214}{\rm Bi}$ events can be tagged with high efficiency by the subsequent $\alpha$-decay 
of $^{214}{\rm Po}$. It was, e.g. noted in \cite{Biller} that 
$99.8\%$ of the decay of $^{214}{\rm Po}$ occur outside trigger windows and 
thus can be efficiently identified in SNO+. In the remaining $0.2\%$ cases, 
a discrimination can be obtained by looking at the time structure of the 
generated signal. 
All this shows that the $^{214}{\rm Bi}$ background 
can be potentially reduced at a level of $\sim {\rm few}\,{\rm counts}/{\rm year}/ {\rm kton}$ 
or less (integrated over the entire energy spectrum) thus allowing for the proposed measure.

 In the $^{232}{\rm Th}$ chain, the potentially dangerous radioisotope is 
$^{208}{\rm Tl}$ which undergoes $\beta$ decay to excited states of $^{208}{\rm Pb}$ 
followed by $\gamma$ particles emitted in the transitions to the $^{208}{\rm Pb}$-ground state.
The total energy of the emitted particles is equal to $Q=5.0\,{\rm MeV}$
with a minimum energy released in the detector equal to $\sim 2.6\, {\rm MeV}$
that corresponds to
the transition from first excited state to ground state of $^{208}{\rm Pb}$. 
The $^{208}{\rm Tl}$ background, thus, fall outside the energy window proposed for the 
identification of ecCNO neutrinos. This is also shown in \cite{Lena8B} where the event rate 
expected in the LENA detector between $[1.5,3.0]\;{\rm MeV}$ is calculated\footnote{
We note, for completeness, that $^{232}{\rm Th}$ contamination in Borexino is $(3.8\pm0.8)\times 10^{-18} \,{\rm g/g}$.
This corresponds to a total rate $R(^{208}{\rm Tl})\sim 5\times 10^2 {\rm counts}/{\rm year}/ {\rm kton}$.  
The amount of $^{208}{\rm Tl}$ can be determined from the observed number of 
$^{212}{\rm Bi}- \,^{212}{\rm Po}$ coincidences (see e.g. \cite{BorexinoLong}).}.
 
The cosmogenic background is due to muon-induced
production of radioactive nuclides. The majority of the 
produced radioisotopes have a short lifetime and, thus, 
the associated background can be efficiently rejected 
by vetoing the detector few seconds after the muon 
passage. The remaining cosmogenic isotopes with long
lifetimes are $^{10}{\rm C}$, $^{11}{\rm Be}$ 
and  $^{11}{\rm C}$. The production rate of these nuclei 
is roughly 
proportional to the muon flux at the experimental site
which is equal to $\Phi_{\mu}=28.8 \;{\rm m}^{-2}\,{\rm d}^{-1}$ for LNGS (Borexino),
$\Phi_\mu= 4.8 \;{\rm m}^{-2}\,{\rm d}^{-1}$ for Pyh\"{a}salmi (LENA) and $\Phi_{\mu} = 0.288 \;{\rm m}^{-2}\,{\rm d}^{-1}$
for SNOLAB (SNO+). 
Following \cite{Lena8B}, we estimate the cosmogenic background in different detectors 
by rescaling the Borexino rates~\cite{BorexinoLong} proportionally to $\Phi_\mu$,
obtaining the results given in tab.~\ref{tab3}.

\begin{table}[t]
\begin{center}{
\begin{tabular}{l|cc|c|ccc|ccc}
 Isotope   &   $\tau $       & $Q({\rm MeV})$   & $R^{\rm tot}_{\rm LNGS} $ &       & $R^{\rm tot}_{\rm Pyh} $ &        &             & $R^{\rm tot}_{\rm SNO}$ &      \\
\hline
$^{10}{\rm C}$   & 27.8 s     & 3.7    &  $1970$ & $330$ & $\to$                      & $6.0$   &  20   & $\to $                    & 0.36   \\
$^{10}{\rm Be}$  & 19.9 s     & 11.5  & $128$                       & $21$   & $\to$                      & $0.08$ &  1.3  & $\to $                    & $0.005$  \\
$^{11}{\rm C}$   & 29.4 min & 2.0    & $1.04 \times 10^5$  &          &   $17 \times 10^3$ &             &         & $1.0 \times 10^3$ &   \\
\hline
\end{tabular}
}\end{center}\vspace{0.4cm} \caption{\em {\protect\small The background rates
produced by long-lived cosmogenic radio-isotopes for a detector located in LNGS, Pyh\"{a}salmi and SNOLAB. 
The rates are expressed in ${\rm counts/year/kton}$ and are integrated over the entire energy spectrum. 
The two numbers that are shown for $^{10}{\rm C}$ and $^{11}{\rm Be}$ nuclei are the rates expected before
(left)  and after (right) introducing a veto for a cylinder with $2 {\rm m}$ radius around each traversing muon for a time 
$\Delta t = 4\cdot\tau(^{10}{\rm C}) = 111.2\, {\rm s}$.
\label{tab3} 
}}\vspace{0.4cm}
\end{table}

As it is discussed in \cite{Lena8B}, since $^{10}{\rm C}$
and $^{11}{\rm Be}$ have a much shorter lifetime than $^{11}{\rm C}$,
the background from these isotopes can be reduced
by vetoing a cylinder with $2 {\rm m}$ radius around each 
traversing muon for a time $\Delta t = 4\cdot\tau(^{10}{\rm C}) = 111.2\, {\rm s}$.
The suppression factor of the $^{10}{\rm C}$ and $^{11}{\rm Be}$ rates 
are approximately equal to $\exp\left(-\Delta t/\tau(^{10}{\rm C})\right) \simeq 2\times 10^{-2}$ and 
$\exp\left(-\Delta t/\tau(^{11}{\rm Be})\right) \simeq  4\times 10^{-3}$ with 
an introduced dead time equal to about 10\% of the total exposure
in Pyh\"{a}salmi and less than $1\%$ in SNOLAB. The resulting background rates 
are given in the last two columns of Tab.~\ref{tab3} from which we see 
that cosmogenic production of $^{10}{\rm C}$ and $^{11}{\rm Be}$ nuclei
do not prevent ecCNO neutrino detection if the detector 
is as deep as LENA or SNO+\footnote{In this work we do not discuss 
the potential of JUNO~\cite{JUNO}. Indeed, this experiment, being closer to the surface, 
is affected by a larger cosmogenic background that cannot be vetoed as discussed above.}. 

\begin{figure}[t]
\par
\begin{center}
\includegraphics[width=8cm,angle=0]{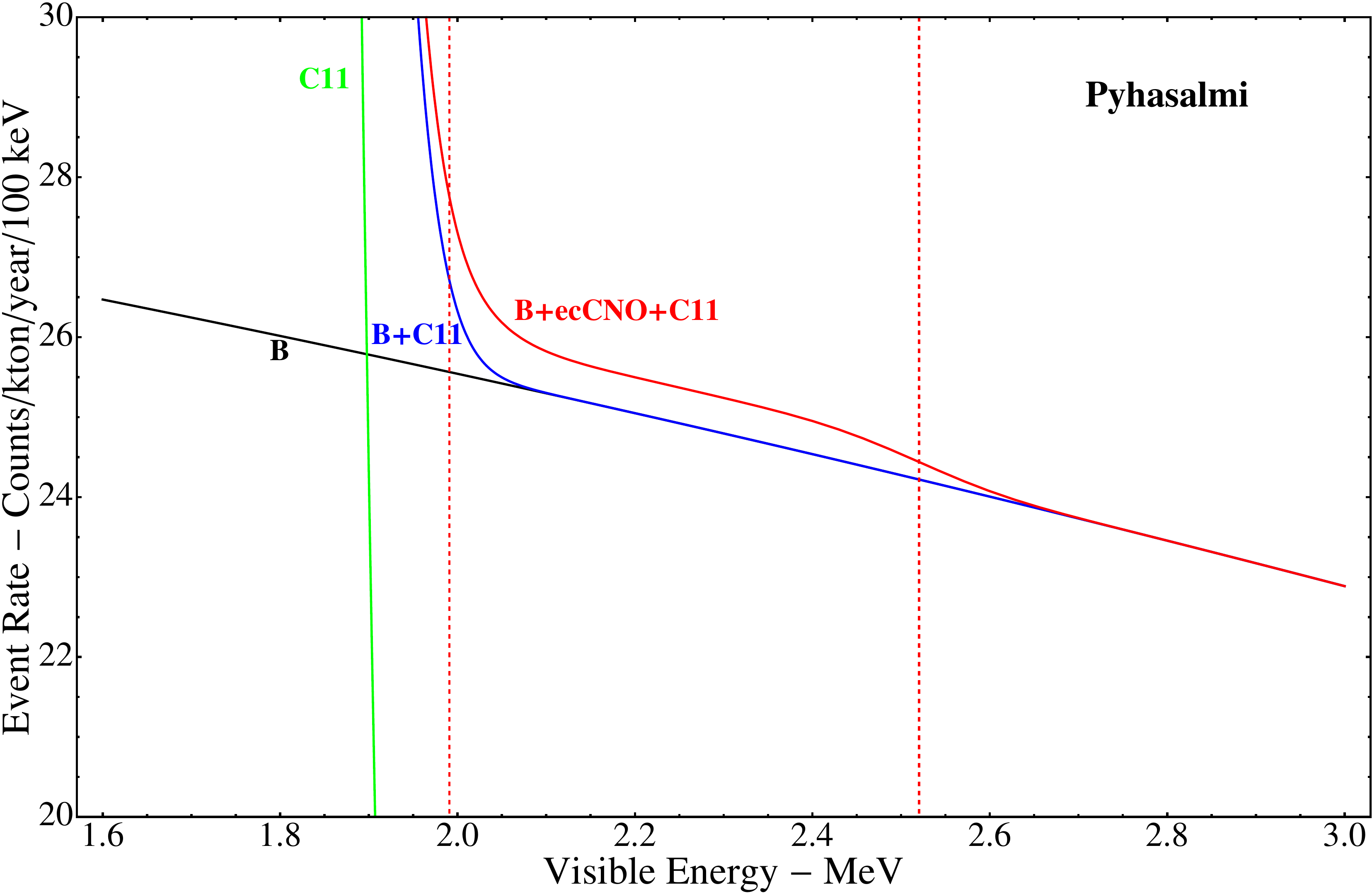}
\includegraphics[width=8cm,angle=0]{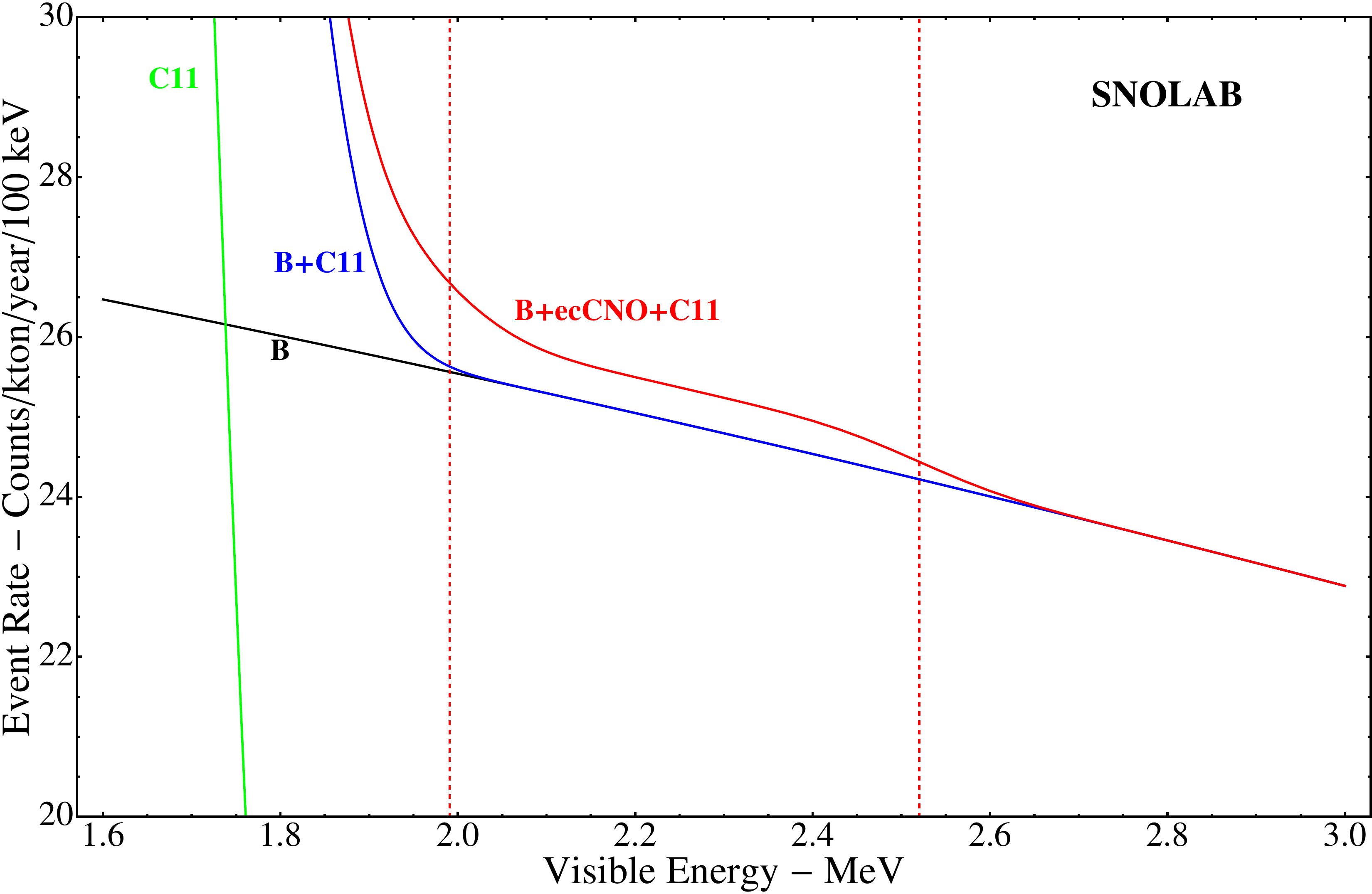}
\end{center}
\par
\vspace{-0.5cm}
\caption{\em\protect\small The expected event rate as a function of the visible energy $E_{\rm vis}$ for
a liquid scintillator detector located at Pyh\"{a}salmi (left) and SNOLAB (right).}
\vspace{0.5cm} 
\label{Fig2}
\end{figure}

 The $^{11}{\rm C}$ cosmogenic background has a much larger rate 
and partially overlap with the energy window considered for the 
proposed measure. In fact, $^{11}{\rm C}$ nuclei undergo $\beta^+$ decay producing 
a positron with a continuous energy spectrum ($Q=1.98\,{\rm MeV}$) 
which subsequently annihilates in the detector producing two gammas 
with $E_{\gamma}=m_{\rm e} = 0.511\,{\rm MeV}$.
The visible energy $E_{\rm vis}$ produced in the detector is calculated by  
using $E_{\rm vis} = E_{e} + k \cdot 2 m_{e}$, where $E_{e}$ is the positron energy 
and the factor $k=0.89$ takes into account that
the scintillation light emitted when a $\gamma$ particle 
is fully absorbed is significantly lower than the light emitted 
by a $\beta$ particle with the same energy.
The adopted value for the parameter $k$ has been estimated 
by comparing the quenching factors of electrons and gammas 
in Borexino as they are deduced from Fig.~8 and Fig.~46 of \cite{BorexinoLong}\footnote{Note that 
Borexino active medium is a solution of PPO and pseudocumene, see \cite{BorexinoLong} for details. 
We assume that the deduced value of $k$ is adequate also for LAB.}.
The obtained spectrum is then convolved with the assumed detector
energy resolution obtaining the results which are presented in Fig.~\ref{Fig2} 
for a detector in  Pyh\"{a}salmi (left panel) and SNOLAB (right panel). 
The cosmogenic $^{11}{\rm C}$ background rates in the energy
windows $[E_{\rm low},2.5 \,{\rm MeV}]$ are given in tab.~\ref{tab4}
as a function of $E_{\rm low}$.
We see that they are comparable with (lower than) the irreducible background produced 
in the same energy range by $^8{\rm B}$ neutrinos for $E_{\rm low} \ge 1.8\,{\rm MeV}$ 
($E_{\rm low} \ge 1.5\,{\rm MeV}$) 
in Pyh\"{a}salmi (in SNOLAB).

 We use the above numbers to estimate of the significativity $\Sigma_{\rm i}$
of a possible ecCNO neutrino measurement in Pyh\"{a}salmi or SNOLAB. As done in the previous section, 
we compare the expected signal with the statistical fluctuations of the total background:
\begin{equation}
\Sigma_{\rm i} = \frac{{\mathcal S}}{\mathcal B} = 
\frac{\left(R_{\rm eN}+R_{\rm eO}\right)}{\sqrt{R_{\rm B}+R_{\rm i}(^{11}{\rm C}) }} \, \sqrt{\mathcal E} 
\label{Sigmai}
\end{equation}
where $i={\rm Pyh},\,{\rm SNO}$ and we included the $^{11}{\rm C}$ contribution (other background 
sources are not considered assuming that their rates are reduced at a level 
much lower than $R_{\rm B}$). The quantity ${\mathcal E}$ indicates the detector exposure and
should be calculated by including the small dead-time ($\sim 10\%$ in Pyh\"{a}salmi and $\le 1\%$ in SNOLAB) 
introduced by cosmogenic cuts. We obtain:
\begin{equation}
\nonumber
\Sigma_{\rm i}   =  \eta_{\rm i}  \, \sqrt{\mathcal E / ({\rm10  kton}\times{\rm year})} 
\label{eta}
\end{equation}
with the parameters $\eta_{\rm Pyh}$ and $\eta_{\rm SNO}$ 
given in the two right columns of tab.~\ref{tab4}. We see that $\eta_{\rm Pyh}\sim 1$ for $E_{\rm low}\simeq 1.8\,{\rm MeV}$
and $\eta_{\rm SNO}\ge 1$ for $E_{\rm low}\le 1.9\,{\rm MeV}$ indicating that the proposed measure,  
despite being extremely difficult, is not excluded from the statistical point of view.
According to our estimate, a $20\;{\rm kton}$ detector located in Pyh\"{a}salmi 
collects a sufficient number of events ($\sim 100 \, {\rm counts}/{\rm year}$ above 1.8~MeV from ecCNO neutrinos) 
to extract the signal with a statistical significance of $\sim 3\sigma$ in $5\,{\rm years}$ of data taking.
The significativity of the extraction could increase to $\sim 3.8 \sigma$ for a detector with the same
characteristics placed at SNOLAB.

\begin{table}[t]
\begin{center}{
\begin{tabular}{l|cc| cc}
 $E_{\rm low} ({\rm MeV})$  &   $R_{\rm Pyh}(^{11}{\rm C}) $       & $R_{\rm SNO}(^{11}{\rm C}) $    &  $\eta_{\rm Phy}$      & $\eta_{\rm SNO}$    \\
\hline 
1.5 & 3130 & 187  &  0.6 &   1.6 \\
1.6 & 1470 & 88.4 &  0.7  &   1.5 \\
1.7 & 500   &  30.0 &  0.8 &   1.5 \\
1.8 & 98.0  & 5.9 &   1.0 &   1.2 \\
1.9 &  7.8   &  0.5  &  0.9  &   1.0 \\
2.0 &  0.2   &  0.0  &  0.7  &   0.7 \\
\hline
\end{tabular}
}\end{center}\vspace{0.4cm} \caption{\em {\protect\small 
The background rate (${\rm counts/year/kton}$) produced by $^{11}{\rm C}$
in the visible energy window $[E_{\rm low},\, 2.5 {\rm MeV}]$ 
for a detector located in Pyh\"{a}salmi and SNOLAB. The last two columns
give the predicted values of the sensitivity parameters $\eta_{\rm Phy}$
and $\eta_{\rm SNO}$ defined in eq.~(\ref{Sigmai}).
\label{tab4}
}}\vspace{0.4cm}
\end{table}

\section{Conclusions}
\label{summary}

 In this work, we analyzed the potential of gigantic ultra-pure
liquid scintillator detectors for the detection of ecCNO neutrinos. 
The obtained results are encouraging as indicated by the fact that 
the sensitivity parameters $\eta_{\rm Pyh}$ and $\eta_{\rm SNO}$ (defined in 
eq.~(\ref{eta})) that 
give the statistical significance of a measure with an exposure 
of ${\rm10  kton}\, \times{\rm year}$ in Pyh\"{a}salmi and SNOLAB,
are $\sim 1$.
Few comments are necessary to further elaborate on our results:\\
{\em i)} Below $2.5\;{\rm MeV}$, ecCNO neutrinos provide a contribution
to the total signal that is comparable to the statistical fluctuations 
for a detector with an exposure ${\mathcal E} = 10\,{\rm kton}\times{\rm year}$ or larger. 
This means that they cannot be neglected in statistical analysis that aim at the
reconstruction of the low energy upturn of the electron neutrino survival probability predicted 
by the LMA-MSW solution of the solar
neutrino problem;\\
{\em ii)} According to our estimate, the detection of ecCNO neutrinos 
in the proposed 50~kton LENA detector cannot be excluded. 
In a recent study \cite{Lena8B}, it was shown that 
the external background in LENA can be reduced to a negligible level 
in a fiducial mass of 19~kton, thus allowing to measure the 
$^{8}{\rm B}$ solar neutrinos event spectrum down to 
$E_{\rm vis} \sim 1.9\,{\rm MeV}$ and to explore the energy region where 
the contribution of ecCNO neutrinos is not negligible. 
Our background estimates are derived along the same lines of \cite{Lena8B} 
and agree with this analysis.
However, that it would advisable that the 
LENA experimental collaboration investigates the actual possibility
of observing ecCNO neutrinos with a complete detector simulation 
and optimized cuts.\\
{\em iii)} In order to go beyond detection and to use ecCNO neutrinos as a probe for the solar composition 
and/or to observe the low energy upturn of $P_{\rm ee}$,  an accuracy at the level of $\sim 15\%$ or better is required. 
Indeed, the predictions for the ecCNO neutrino fluxes disagree by $\sim 30\%$ 
when different surface compositions are considered. 
Incidentally, the electron neutrino survival probability at ecCNO neutrino energies
is also $\sim 30\%$ larger than the high energy value.  
In an ideal detector, with the characteristic described in our analysis 
and placed so deep that the cosmogenic background is negligible, 
the $15\%$ accuracy goal corresponds to an exposure ${\mathcal E} \ge 100 \; {\rm kton} \times {\rm year}$.\\
{\em iv)} Finally, our results are based on a simplified description of the detector 
properties. Being the signal extremely small, the results may critically depend on 
the assumed detector characteristics, e.g. the assumed purification levels; the parametrization of the 
energy resolution function; the description of the detector response (i.e. the $k$ parameter).
As an example, the parameter $\eta_{\rm Phy}$ (calculated for $E_{\rm low}=1.8\, {\rm MeV}$) is reduced from $1.0$
to $0.9$ if the detector energy resolution at 1~MeV is increased from $5\%$ to $7\%$ 
and from $1.0$ to $0.8$ if the parameter $k$ is increased from 0.89 to 0.95. 
We look forward for a complete 
analysis by the experimental collaborations working in the field.

\section*{Acknowledgements}
The author thanks G.~Pagliaroli for invaluable collaboration, discussions and suggestions 
and A.~Ianni for very useful comments. 

\newpage

\section*{\sf  References}
\def\refname{

\end{document}